\documentclass[12pt]{article}
\usepackage[dvips]{graphicx}
\usepackage{amssymb}
\usepackage{amsmath}
\usepackage{epsfig}
\usepackage{cite}
\usepackage{hyperref}
\usepackage{bbold}
\usepackage{multirow}

\numberwithin{table}{section}

\def\beq{\begin{equation}}
\def\eeq{\end{equation}}
\def\be{\begin{equation}}
\def\ee{\end{equation}}
\def\bea{\begin{eqnarray}}
\def\eea{\end{eqnarray}}

\DeclareRobustCommand{\SkipTocEntry}[4]{}
\RequirePackage{color}

\begin{document}

\begin{titlepage}
\begin{center}
\rightline{\small }


\vskip 2cm

{\Large \bf Large complex structure flux vacua of IIB\\[0.75em] and the Tadpole Conjecture}
\vskip 1.2cm

Severin L\"ust
\vskip 0.5cm

{\small\it  Jefferson Physical Laboratory, \\
Harvard University, \\
17 Oxford St., \\
Cambridge, MA, 02138, USA.}
\vskip 0.1cm

\vskip 0.8cm

{\tt }

\end{center}

\vskip 1cm

\begin{center}  {\bf Abstract }\\

\end{center}

\noindent
In this note I demonstrate that certain findings on IIB flux compactifications in the large complex structure limit, when combined with recent statistical bounds on the large $h^{2,1}$ limit, are compatible with the Tadpole Conjecture, contrary to previous claims.

\vspace{0.2cm}

\noindent

\vfill

\noindent
September 2021

\end{titlepage}



\subsection*{Introduction}

The fact that compactifications of string theory usually come with a large number of massless moduli fields is a classical problem of string phenomenology.
In many cases this problem gets at least partially resolved by the observation that non-trivial magnetic fluxes along the cycles of the compactification space can generate mass terms for these fields.
However, in the context of IIB and F-theory compactification the universality of this approach was recently challenged by the formulation of the Tadpole Conjecture in \cite{Bena:2020xrh}.
This conjecture suggests a linear relationship between the flux contribution $N_\mathrm{flux}$ to the D3 tadpole 
and the number of complex structure moduli fields $n_\mathrm{moduli}$,
\begin{equation}\label{eq:TC}
N_\mathrm{flux} > \alpha \, n_\mathrm{moduli} \,,\qquad \text{if} \qquad n_\mathrm{moduli} \gg 1 \,,
\end{equation}
where $\alpha$ is a constant of order one.
Since $N_\mathrm{flux}$ is bounded from above by the amount of negative contributions to the tadpole cancellation condition,
depending on the numerical value of $\alpha$
 this conjecture would rule out moduli stabilization at a large number of moduli.
 It would hence challenge previous estimates on the size of the string theory landscape 
 and potentially also common constructions for de Sitter vacua in the context of IIB string theory, such as KKLT or LVS.
 
 Recently, the authors of \cite{Marchesano:2021gyv} provided a detailed analysis of F-theory and IIB flux compactifications at large complex structure and suggested a potential counter example to the above conjecture.
They found two classes of vacua which appear to have qualitatively different behavior with respect to \eqref{eq:TC}.
In both of them all complex structure moduli are stabilized by fluxes.
However, while in the first class the number of flux quanta that contribute positively to $N_\mathrm{flux}$ generically scales linearly with $n_\mathrm{moduli}$,
the second class is designed in such a way that only one pair of flux quanta enters $N_\mathrm{flux}$, independently of $n_\mathrm{moduli}$.
This provokes the conclusion that only the first class of such vacua obeys the tadpole conjecture, while in the second class (the ``IIB2 scenario'') an arbitrarily large number of moduli can be stabilized at only moderately large value of $N_\mathrm{flux}$, seemingly violating \eqref{eq:TC}.

However, in another recent preprint \cite{Plauschinn:2021hkp} the statistical data of \cite{Demirtas:2018akl} on the Kreuzer-Skarke list is used to derive additional bounds on $N_\mathrm{flux}$ in the large complex structure limit of IIB.
The results found there are not only compatible with the tadpole conjecture but suggest an even more dramatic, polynomial scaling behavior of $N_\mathrm{flux}$ with $n_\mathrm{moduli}$.
Yet, the analysis in \cite{Plauschinn:2021hkp} differs from the one in \cite{Marchesano:2021gyv} as it does not include certain correction terms which were crucial for the construction of the potential counter example of \eqref{eq:TC}.

In this note we would like to demonstrate that taking the estimates of \cite{Plauschinn:2021hkp} at face value and combining them with the results of \cite{Marchesano:2021gyv} on IIB vacua indicates that also in the 
second class of vacua in \cite{Marchesano:2021gyv} 
$N_\mathrm{flux}$ cannot be made arbitrarily small and scales approximately linear with $n_\mathrm{moduli}$, in agreement with \eqref{eq:TC}.

We proceed by summarizing a few elementary facts on flux compactifications in the large complex structure limit of IIB. Afterwards, we collect the relevant results of \cite{Marchesano:2021gyv} and \cite{Plauschinn:2021hkp} and show that when combined they support the tadpole conjecture, even when taking polynomial corrections to the prepotential into account.

\subsection*{IIB flux compactifications at large complex structure}

In order to keep this note as short as possible we will only introduce a minimal set of concepts which are needed to setup our notation  and do not aim at being self-contained.
For more details we refer the reader to the vast body of existing literature on this topic.

IIB string theory on an orientifold of a Calabi-Yau three-fold $X$ allows for non-trivial fluxes of the three-form field strengths $F_3$ and $H_3$ along its three-cycles.
By expanding $F_3$ and $H_3$ in terms of a symplectic basis $(\alpha_I, \beta^J)$, $I,J = 0, \dots, h^{2,1}$, of $H^3(X, \mathbb{Z})$ we denote the individual integer flux quanta by
\begin{equation}
F_3 = f^I \alpha_I - f_I \beta^I \qquad \text{and} \qquad H_3 = h^I \alpha_I - h_I \beta^I \,.
\end{equation}
Such fluxes carry a non-trivially D3 charge $N_\mathrm{flux}$ and therefore contribute to the D3 tadpole cancellation condition according to
\begin{equation}
2 N_\mathrm{flux} = \int_X F_3 \wedge H_3 = h^I f_I - h_I f^I \,.
\end{equation}
A self-duality condition relates $F_3$ and $\star H_3$ and enforces $N_\mathrm{flux} > 0$.
On the other hand, $N_\mathrm{flux}$ is bounded from above by the amount of negative D3 charge carried by O3- and O7-planes as well as D7-branes.

The fluxes generically generate a scalar potential for the axio-dilaton $\tau$ and the complex structure moduli $z^i$, $i = 1, \dots, h^{2,1}$, and are hence able to stabilize them at specific values.
Therefore, we have $n_\mathrm{moduli} = h^{2,1} + 1$.
All information on the geometry of the corresponding moduli space and the flux-induced potential can be encoded in terms of the prepotential $\mathcal{F}(z^i)$.
At large complex structure, including possible polynomial corrections, it reads
\begin{equation}\label{eq:prepotential}
\mathcal{F} = -\frac{1}{6} \kappa_{ijk} z^i z^j z^k + \frac12 K_{ij}^{(1)} z^i z^j + K_i^{(2)} z^i - \frac{i}2 K^{(3)} \,,
\end{equation}
where $\kappa_{ijk}$ are the triple intersection numbers on $H^{2}(\widetilde{X}, \mathbb{Z})$ of the mirror dual three-fold $\widetilde{X}$.
While $K_{ij}^{(1)}$ and $K_i^{(2)}$ are not of particular relevance for the following discussion, it is worth noting that $K^{(3)}$ is given in terms of the Euler number $\chi$ of $X$ as
\begin{equation}\label{eq:euler}
K^{(3)} = \frac{\zeta(3) \chi}{(2 \pi)^3} = \frac{\zeta(3)}{4 \pi^3} \left(h^{1,1} - h^{2,1} \right) \,,
\end{equation}
where in the last step we expressed $\chi$ in terms of the Hodge numbers.

It is often convenient to split the complex structure moduli in their real and imaginary parts,
\begin{equation}
z^i = b^i + i \, t^i \,,
\end{equation}
usually called axions and saxions, respectively.
For future reference we also introduce the contractions
\begin{equation}
\kappa_i = \kappa_{ijk} t^jt^k \,,\qquad\qquad \kappa = \kappa_i t^i \,,
\end{equation}
where the $\kappa_i$ are sometimes called dual saxions.
Importantly, with a prepotential of the form \eqref{eq:prepotential}, the moduli space metric $g_{ij}$ depends only on the saxions $t^i$ and is independent of the axions $b^i$.
Consequently, not all values of the saxions $t^i$ correspond to points in the moduli space of Calabi-Yau metrics.
Neglecting the correction terms, e.g.~for  $\kappa \gg \left | K^{(3)} \right |$, 
this space of admissible values for the $t^i$ is given by the K\"ahler cone of the mirror dual Calabi-Yau $\tilde X$.
For smaller values of $t^i$, however, when the corrections cannot be ignored, this space gets deformed and is not a cone anymore.

\subsection*{The IIB2 scenario of \cite{Marchesano:2021gyv}}

This setup considers fluxes where the only non-vanishing flux quanta are given by
\begin{equation}
(f^I, f_J, h_0) \,,
\end{equation}
i.e.~$h_i = h^I = 0$.
Consequently, the contribution to the tadpole cancellation condition reads
\begin{equation}
N_\mathrm{flux} = - \frac12 h_0 f^0 \,,
\end{equation}
and contains only one pair of flux quanta.
It was argued in \cite{Marchesano:2021gyv} that by choosing suitable values for the remaining fluxes $f^i$ and $f_I$ which do not enter $N_\mathrm{flux}$, all moduli, including both axions and saxions, can be stabilized.
Therefore, $N_\mathrm{flux}$ appears to be independent of the number of moduli and \eqref{eq:TC} seems to be violated.

Importantly, the stabilization of the saxions depends crucially on the presence of the correction $K^{(3)}$ in the prepotential \eqref{eq:prepotential}.
In particular,
\cite{Marchesano:2021gyv} derived the following bound on $N_\mathrm{flux}$ in terms of $K^{(3)}$ and the (dual) saxion vevs,
\begin{equation}\label{eq:MPWbound}
N_\mathrm{flux} \gtrsim d \left( \frac{1}{K^{(3)}} \frac{\kappa}{\kappa_i}\right)^\frac12 \,,
\end{equation}
where $d = \mathrm{gcd}(f^0, h_0)$.  
We would now like to show that using the estimates which were made in \cite{Plauschinn:2021hkp} this bounds scales almost linearly with the number of moduli.

The most conservative bound can be obtained by assuming that $d$ does not scale with the number of moduli.
We will therefore from now on assume $d = \mathcal{O}(1)$ and drop it together with all other order one constants from the following estimates.

\subsection*{The estimates of \cite{Plauschinn:2021hkp}}

In \cite{Demirtas:2018akl} topological properties of a large number of Calabi-Yau threefolds in the Kreuzer-Skarke list were statistically analyzed, focusing on the large $h^{1,1}$ regime.
Using mirror symmetry these results were translated in \cite{Plauschinn:2021hkp} into bounds on varies quantities on the complex structure moduli space of the dual three-folds in the large complex structure and large $h^{2,1}$ limit.
The first observation which was made there is that the number of non-vanishing components of the triple intersection numbers $\kappa_{ijk}$ scales roughly linearly with the number of moduli, i.e.
\begin{equation}
\# \left(\kappa_{ijk} \neq 0 \right) \gtrsim h^{2,1} \,,
\end{equation}
where we neglected an order one proportionality constant and a constant term.
Using the fact that the average component of a generic vector scales like $v^i \sim d^{-1/2} \|v\|$ with its dimension $d$ and its Euclidean norm $\|v\|$,
\cite{Plauschinn:2021hkp} arrives at the estimate
\begin{equation}\label{eq:kappabound}
\kappa \gtrsim \left(h^{2,1}\right)^{-\frac12}  \left\| t \right \|^3 \,.
\end{equation}
Along the same lines, keeping in mind that the vector $\kappa_i$ has generically $h^{2,1}$ non-vanishing components,%
\footnote{I would like to thank Erik Plauschinn for helpful discussion on this point.}
 we obtain\begin{equation}
\kappa_i \gtrsim \left(h^{2,1}\right)^{-1} \left\| t \right \|^2 \,,
\end{equation}
and hence
\begin{equation}\label{eq:kappakappabound}
\frac{\kappa}{\kappa_i} \gtrsim \left(h^{2,1}\right)^{\frac12}  \left\| t \right \| \,.
\end{equation}

Importantly, \cite{Plauschinn:2021hkp} also uses the results of \cite{Demirtas:2018akl} to argue that $\left\| t \right \| $ has to scale with a positive power of $h^{2,1}$ if one wants to maintain control over the corrections in the large complex structure prepotential \eqref{eq:prepotential}.
For this purpose \cite{Demirtas:2018akl} introduces the concept of the so-called stretched K\"ahler cone which consists of all points in the K\"ahler cone that have a minimal distance $c$ to its boundaries.
Moreover, they found that the smallest angle between two boundaries of the K\"ahler cone scales inversely with the number of moduli, the K\"ahler cone becomes narrower for larger $h^{2,1}$.
As illustrated in Figure~1 of \cite{Plauschinn:2021hkp} this implies that the minimal distance $d_\mathrm{min}$ between the apex of the K\"ahler cone and a point within the stretched K\"ahler cone increases with $h^{2,1}$.
Concretely, a log-log-linear fit against the data of \cite{Demirtas:2018akl} yields
\begin{equation}\label{eq:dmin}
d_\mathrm{min} \sim c \left(h^{2,1}\right)^\beta \,,
\end{equation}
with $\beta \approx 2.5$ (in \cite{Plauschinn:2021hkp} also the value $\beta \approx 2.7$ for large values of $h^{2,1}$ is quoted), again suppressing the proportionality constant.

\subsection*{Combining the  estimates}

It is now straightforward to combine these estimates and to translate them into a bound on $N_\mathrm{flux}$ in the IIB2 scenario of \cite{Marchesano:2021gyv}.
As mentioned above, in the IIB2 scenario full moduli stabilization can only be obtained if the correction term $K^{(3)}$ is taken into account.
On the other hand, \cite{Plauschinn:2021hkp} studies moduli stabilization in a regime where all correction terms can be neglected.
This motivates the introduction of the (mirror dual of) the stretched K\"ahler cone, which also could be called the stretched complex structure cone.
All points within this cone have a distance of at least $c$ to its boundaries and hence in its interior all corrections with a magnitude sufficiently smaller than $c$ are suppressed.
This justifies the bound $\|t\| \gg d_\mathrm{min}$.

In the IIB2 case the most significant correction to the large complex structure prepotential is $K^{(3)}$.
We will therefore try  to relate $c$ to $K^{(3)}$.
Even though the correction $K^{(3)}$ must not be neglected, one still has to ensure that its effect is sufficiently small, in order to maintain control over the large complex structure expansion.
Naively, this is the case as long as $\kappa \gg K^{(3)}$.
More importantly, however, the presence of the correction $K^{(3)}$ modifies the metric on the complex structure moduli space in the vicinity of its boundaries and therefore deforms the cone of admissible Calabi-Yau metrics.%
\footnote{It is amusing to think of this deformation as reminiscent of the change in geometry when transitioning from the singular to the deformed conifold in the IR of the Klebanov-Strassler solution \cite{Klebanov:2000hb}.}
Points close to the boundaries of the original complex structure cone, in particular at small $\|t\|$, will not correspond to viable Calabi-Yau metrics anymore and are therefore not part of the moduli space.
The effects of such deformations become more drastic the smaller the angle of the complex structure cone is.
For cones with large angles they are most significant in a regime where $\kappa \approx K^{(3)}$ or using the estimate \eqref{eq:kappabound} for $\kappa$ when
\begin{equation}
\left\| t \right \|^3 \sim \left(h^{2,1}\right)^\frac12 K^{(3)} \,.
\end{equation}
Bringing this back into the language of the stretched complex structure cone leads us to the identification
\begin{equation}\label{eq:cK3}
c \sim \left(h^{2,1}\right)^{1/6} \left(K^{(3)}\right)^{1/3} \sim \left(h^{2,1}\right)^{1/2} \,,
\end{equation}
where in the last step we used  \eqref{eq:euler}.
Therefore, the angle dependence of the stretched complex structure cone translates via \eqref{eq:dmin} into the bound
\begin{equation}
\|t\| \gtrsim d_\mathrm{min} \sim \left(h^{2,1}\right)^{\beta+\frac12} \,.
\end{equation}
Inserting this bound and  \eqref{eq:kappakappabound}  into \eqref{eq:MPWbound} we now readily arrive at
\begin{equation}\label{eq:finalboundA}
N_\mathrm{flux} \gtrsim \left(K^{(3)}\right)^{-\frac12} \left(h^{2,1}\right)^{\frac\beta2+\frac12} \sim \left(h^{2,1}\right)^{\frac\beta2} \,,
\end{equation}
where we used again \eqref{eq:euler} and have suppressed all order one constants.
In particular, for $\beta \approx 2.5$ or $\beta \approx 2.7$ 
this gives a scaling behavior which is slightly stronger than the linear scaling behavior  required by the tadpole conjecture.

We should finally mention that even if one does not relate $c$ directly to $K^{(3)}$ as in \eqref{eq:cK3}, but keeps it as an arbitrary, $h^{2,1}$ independent constant, the bound on $N_\mathrm{flux}$ modifies only marginally and becomes
\begin{equation}\label{eq:finalboundB}
N_\mathrm{flux} \gtrsim c^\frac12 \left(h^{2,1}\right)^{\frac\beta2-\frac14} \,,
\end{equation}
which for  $\beta \approx 2.5$ or $\beta \approx 2.7$ gives $N_\mathrm{flux} \gtrsim \left(h^{2,1}\right)^{1.0}$ or $N_\mathrm{flux} \gtrsim \left(h^{2,1}\right)^{1.1}$, respectively.

Even though we did not attempt to collect all numerical order one coefficients which go into the constant $\alpha$ in \eqref{eq:TC} and even though the exact power in \eqref{eq:finalboundA} or \eqref{eq:finalboundB} depends on relatively rough estimates, we can conclude that $N_\mathrm{flux}$ of the IIB2 scenario scales non-trivially with $h^{2,1}$, in agreement with the main message of the tadpole conjecture.%
\footnote{While completing this note it was brought to my attention that a similar observation was independently made by the authors of 
an upcoming paper 
\cite{GPH}.
I would like to thank Thomas Grimm, Erik Plauschinn and Damian van de Heisteeg for correspondence on their work.
}

Notably, \cite{Marchesano:2021gyv} studies mainly full F-theory setups and not only the IIB limit.
It proposes an analogous scenario for moduli stabilization in F-theory where $N_\mathrm{flux}$ depends only on one pair of fluxes.
It would be very interesting to see if similar considerations can be used there to detect a possible ``hidden'' scaling behavior with the number of moduli.%

\vskip 1cm
\noindent {\bf Acknowledgements:} 

I would like to thank Iosif Bena, Mariana Gra\~na, Thomas Grimm, Fernando Marchesano, Dieter L\"ust and Erik Plauschinn for very helpful discussions and correspondence.
This work was supported by the National Science Foundation grant NSF PHY-1915071.


\hskip 1cm


\providecommand{\href}[2]{#2}\begingroup\raggedright\endgroup


\begin{thebibliography}{10}

\bibitem{Bena:2020xrh}
I.~Bena, J.~Bl\r{a}b\"ack, M.~Gra\~na and S.~L\"ust,
``The Tadpole Problem,''
[arXiv:2010.10519 [hep-th]].

\bibitem{Marchesano:2021gyv}
F.~Marchesano, D.~Prieto and M.~Wiesner,
``F-theory flux vacua at large complex structure,''
JHEP \textbf{08} (2021), 077
[arXiv:2105.09326 [hep-th]].

\bibitem{Plauschinn:2021hkp}
E.~Plauschinn,
``The tadpole conjecture at large complex-structure,''
[arXiv:2109.00029 [hep-th]].
        
\bibitem{Demirtas:2018akl}
M.~Demirtas, C.~Long, L.~McAllister and M.~Stillman,
``The Kreuzer-Skarke Axiverse,''
JHEP \textbf{04} (2020), 138
[arXiv:1808.01282 [hep-th]].
        
\bibitem{Klebanov:2000hb}
I.~R.~Klebanov and M.~J.~Strassler,
``Supergravity and a confining gauge theory: Duality cascades and chi SB resolution of naked singularities,''
JHEP \textbf{08} (2000), 052
[arXiv:hep-th/0007191 [hep-th]].

\bibitem{GPH}
T.~W.~Grimm, E.~Plauschinn, D.~van de Heisteeg,
``Moduli Stabilization in Asymptotic Flux Compactifications,''
to appear.
        
\end{thebibliography}
\end{document}